\documentstyle[aps,epsf,multicol,prl]{revtex}

\input epsf
\begin{document}
\draft
\title{Unstable decay and state selection}  
\author{Martin B. Tarlie$^{1}$ and Alan J. McKane$^{2}$}
\address{$^{1}$James Franck Institute, University of Chicago, 
5640 South Ellis Avenue, Chicago, IL 60637 \\
$^{2}$Department of Theoretical Physics,
University of Manchester, Manchester M13 9PL, UK }      
\maketitle      
\begin{abstract}
We consider the problem of state selection for a stochastic system, initially
in an unstable stationary state, when multiple metastable states compete for
occupation. Using path-integral techniques we derive 
remarkably simple and accurate formulas for state-selection
probabilities. The method is sufficiently general that it is applicable to
a wide variety of problems.
\end{abstract}
\pacs{PACS numbers: 05.40.+j, 02.50.Ey, 05.20.-y} 

\begin{multicols}{2}

The investigation of the decay from a metastable state has been the subject
of numerous studies over very many years 
\cite{ref:kramers,ref:risken,ref:gardiner}. 
But the analogous problem
of the decay from an unstable state has received comparatively little 
attention \cite{ref:gardiner,ref:suzuki,ref:caroli,ref:weiss}, 
and what studies 
there have been have focussed on the kinetic properties of one-dimensional and
quasi-one-dimensional systems \cite{ref:quasi}. However, these studies cannot
address one of the fundamental, open questions in non-equilibrium statistical
mechanics: state selection from an unstable state in systems 
with multiple, isolated minima. Here 
we present a systematic, intuitive, and analytically tractable
method which gives results in excellent agreement with Monte-Carlo simulations.

When driven far from equilibrium many systems encounter instabilities. At such
points, noise plays a crucial role. In addition, in complex systems there are 
multiple modes that interact and can compete. Perhaps the most familiar 
example is found in Rayleigh-B\'enard convection. 
Consideration of the interaction between two competing modes leads
to the following equations for their amplitudes $x$ and $y$ \cite{ref:segel}:
\begin{eqnarray}
\dot x &=& \alpha x - \gamma x y^{2} - \delta x^{3} +\eta_{x}(t)\cr
\dot y &=& \beta y - \gamma y x^2 - \epsilon y^{3} +\eta_{y}(t),
\label{eq:langevin}
\end{eqnarray}
where $\alpha$ and $\beta$ are the (positive) growth rates for the two modes 
$x$ and $y$, $\gamma$ is the (positive) coupling coefficient, and 
$\delta$ and $\epsilon$ are positive stabilizing coefficients. 
The variables $\eta_{x}$ and $\eta_{y}$ 
are Gaussian random variables with mean zero and 
variance $\langle \eta_{i}(t)\eta_{j}(t')\rangle=2 D\delta_{ij}\delta(t-t')$, 
where $i$ and $j$ are either $x$ or $y$, and $D$ is the noise strength. 
As we are considering the decay from the
unstable stationary point $x\!=\!0,y\!=\!0$, the noise plays an essential role.
There are four main elements that are present in Eq.~(\ref{eq:langevin}): 
(i) an unstable stationary point with exponential growth of the modes in the
neighborhood of this point, (ii) interaction between the modes,  
(iii) isolated metastable states, and (iv) noise. 
These features are also found in many other systems 
\cite{ref:cross,ref:kramer,ref:bagchi}.  

In this Letter we address the question: given a system described by equations 
such as (\ref{eq:langevin}), with the initial condition being the 
unstable stationary point,
what is the probability that the system finds itself in a given metastable
configuration? The system will relax to thermal equilibrium over a time 
scale that is on the order of $\exp (E/D)$, where $E$ is a 
characteristic energy barrier separating the metastable states. However, 
if $D\ll 1$, this time can be enormous. Our focus is on understanding the 
occupation over shorter time scales. 

Our approach is based on the path-integral representation for the conditional
probability density $P({\bf r},T|{\bf 0},0)$ that the system resides in state
${\bf r}$ at time $T$ given that it started at the unstable stationary point
at the origin. For purposes of illustration, we take the concrete, 
physically important example presented in Eq.~(\ref{eq:langevin}). 
The path-integral expression for $P$ is given by \cite{ref:graham}
\begin{eqnarray}
P&=&\int {\cal D}{\bf r}\, J[{\bf r}]\,e^{-S[{\bf r}]/D}
\label{eq:pathint}
\\
{\rm where \ \ \ \ }
S[{\bf r}]&=&\frac{1}{4}\int_{0}^{T}\!dt\, [\dot {\bf r}+
\nabla V({\bf r}) ]^{2} 
\label{eq:action}
\\
{\rm and \ \ \ \ \ }J[{\rm r}]&=&\exp \Big( \frac{1}{2}\int_{0}^{T}\!dt\,
\nabla^{2}V({\bf r})\Big).
\label{eq:jacobian}
\end{eqnarray}
Here $S$ is the action, $J$ is the Jacobian, and $V$ 
is the potential for this problem and is given by 
\begin{equation}
V({\bf r})=-\frac{\alpha}{2} x^2 - \frac{\beta}{2} y^2 + 
\frac{\gamma}{2} x^2 y^2 + \frac{\delta}{4} x^4 + \frac{\epsilon}{4} y^4.
\label{eq:V}
\end{equation} 

To evaluate the path integral for weak noise, a natural approximation scheme 
is the method of steepest descent. In this 
approach, the path integral is dominated by the paths of least action;
a necessary condition is that these paths make the action stationary. 
The leading approximation is to simply evaluate the action along these 
paths. However, in our case it is necessary to go beyond this order and 
include both (Gaussian) fluctuations about the paths of least action, 
as well as the Jacobian evaluated along the appropriate path. 
In other words, once the stationary paths have been determined, 
we need to calculate three quantities: (i) the action, (ii) the 
Jacobian, and (iii) the fluctuation determinant, which characterizes the 
effect of fluctuations about the relevant path.   

\begin{figure}
\narrowtext
\epsfxsize=\hsize 
\epsfysize=3.4in
\epsfbox{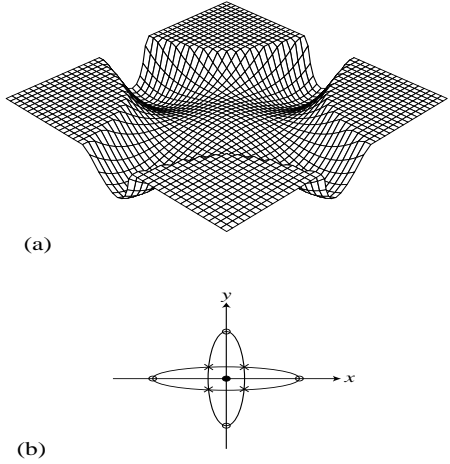}
\vspace{0.1in}
\caption{(a) $V(x,y)$ with $\alpha = \beta = 2$, $\gamma =4$, 
$\delta = \epsilon =1/5$. (b) Contours of zero force.}
\label{fig:V}
\end{figure}

Having outlined a general prescription for calculating
the conditional probability density, we now focus on the specific 
example introduced above. Figure 1a is a three-dimensional plot of 
$V({\bf r})$ for a certain choice of parameters. In Fig. 1b we plot
the locus of points for which $\nabla V=0$. This is a useful way to visualize
state space; the points where the ellipses intersect each other are the 
saddle-points of $V$ (denoted by crosses), the points where the ellipses
intersect the $x$ and $y$ axes are the local minima (denoted by open circles),
and the origin is the unstable stationary point (denoted by a closed circle).
The question of interest here can now be phrased in the following way:
given an ensemble of systems, each of which starts at the unstable stationary
point, what fraction of the ensemble flows into an $x$-valley or $y$-valley 
(which lead to the $x$- and $y$-wells, respectively)? 
For simplicity, we suppose that $\delta$ and $\epsilon$ are sufficiently
small that the local minima are so distant from the region where 
state selection occurs that they have no influence. Operationally, this 
consists of setting $\delta$ and $\epsilon$ to zero, so that now 
\begin{equation}
V(x,y)=-\frac{\alpha}{2} x^2 - \frac{\beta}{2} y^2 + 
\frac{\gamma}{2} x^2 y^2.
\label{eq:V2}
\end{equation}
Given that the minima are now irrelevant, it is natural to consider the 
conditional probability that if the system starts at ${\bf r}={\bf 0}$ 
it ends up in an $x$-valley denoted by $(X,0)$ or a $y$-valley denoted by
$(0,Y)$. To do this we follow the procedure outlined
above, viz. we first find the path, or paths, of least action that connect the 
unstable point to a point in one of these valleys. (Hereafter we shall confine
our attention to the calculating the probability that if the system starts at
${\bf r}=(0,0)$ at $t=0$ that it end at ${\bf r}_{x}=(X,0)$ at $t=T$. 
The analogous problem where the endpoint is ${\bf r}_{y}=(0,Y)$ can be handled 
in exactly the same way.) The most obvious stationary path is 
${\bf r}_{c}=(x_{c},0)$, where 
\begin{equation}
x_{c}(t)=X\frac{\sinh (\alpha t)}{\sinh (\alpha T)}.
\label{eq:xc}
\end{equation} 
The action $S_{c}\equiv S[x_{c}]$ for this solution is given by 
\begin{equation}
S_{c}=\frac{\alpha X^{2}}{4}[\coth (\alpha T)-1]
\label{eq:Sc1}
\end{equation} 
In the limit that $T\rightarrow \infty$, $S\rightarrow 0$, so that, at 
least in this limit, this solution is a path of least action, not simply
a stationary path. We are unable to prove that this is the only stationary
path that connects $(0,0)$ with $(X,0)$. However, we will show that we can 
make significant progress by considering only this path. In fact, this 
simplification will enable us to 
derive a remarkably simple formula for $P({\bf r}_{x},T|{\bf 0},0)$.

The second factor that we must evaluate is the Jacobian evaluated along the 
path ${\bf r}_{c}$. Using Eqs.~(\ref{eq:jacobian}) and (\ref{eq:xc}) we find
that 
\begin{equation}
J[{\bf r}_{c}]
\equiv J_{x}J_{y}
=
\biggl\{e^{-\frac{1}{2}\int_{0}^{T}\!dt\,\alpha } \biggr\}
\biggl\{e^{ \frac{1}{2}\int_{0}^{T}\!dt\,
        \left[ -\beta +\gamma x_{c}^{2}(t)\right] 
                }\biggr\}.
\label{eq:Jc}
\end{equation}
It is straightforward to calculate both $J_{x}$ and $J_{y}$, with the 
results that 
\begin{eqnarray}
J_{x}&=&\exp \left( -\frac{1}{2}\alpha T \right)
\label{eq:Jx}
\\
{\rm and \ \ \ \ \ }
J_{y}&=&
\exp \left( -\frac{\beta T}{2}+\frac{\gamma X^{2}}{4\alpha} 
\frac{\sinh (2\alpha T)-2\alpha T}{2\sinh^{2} (\alpha T)}
 \right) .
\label{eq:Jy}
\end{eqnarray}

The third quantity that we must calculate is the effect of
fluctuations about ${\bf r}_{c}$. To do this, we expand $S[{\bf r}]$ about
the path ${\bf r}_{c}$, keeping terms of second order. Taking 
${\bf r}={\bf r}_{c}+\delta{\bf r}$, we have that 
$S[{\bf r}]=S[{\bf r}_{c}]+
\frac{1}{2}\int \! dt\,\delta {\bf r}\,L[{\bf r}_{c}]\,\delta {\bf r}$, where
$L_{c}\equiv L[{\bf r}_{c}]$ is a $2\times 2$ matrix-differential operator
that is given by 
\begin{eqnarray}
L[{\bf r}_{c}]&\equiv&
\left[
\matrix{
L_{x}&0\cr
0&L_{y}\cr
}\right]
\\
{\rm where \ \ \ \ }
L_{x}&\equiv& -\partial_{t}^{2}+\alpha^{2}
\label{eq:Lx}
\\
{\rm and \ \ \ \ \ }
L_{y}&\equiv& -\partial_{t}^{2}+
(-\beta +\gamma x_{c}^{2})^{2}-
        2(\alpha x_{c})(\gamma x_{c})
\label{eq:Ly}
\label{eq:Lc}
\end{eqnarray}
The path integral in Eq.~(\ref{eq:pathint}) over ${\bf r}$ now
becomes an integral over $\delta {\bf r}$. Using the second-order 
expansion of $S[{\bf r}]$, the Gaussian integrals over 
$\delta {\bf r}$ can be completed. These integrals contribute a factor
of $\sqrt{|{\rm det}L_{c}|}^{-1}$ to the expression for the conditional 
probability. As $L$ is block-diagonal, we have that 
$\det L_{c}=\det L_{x}\det L_{y}$.  Combining this with 
the fact that $J_{c}=J_{x}J_{y}$, the steepest-descent approximation for 
the conditional probability can be written as
\begin{eqnarray}
P({\bf r}_{x},T|{\bf 0},0)&\sim &\Omega^{-1}P_{0}
\label{eq:Pc1}
\\
{\rm where \ \ }
P_{0}&\equiv&
\frac{J_{x}}{\sqrt{|{\rm det}L_{x}|}}
{\rm e}^{ -S_{c}/D }
\label{eq:P01}
\\
{\rm and  \ \ }
\Omega^{-1}&\equiv& \frac{J_{y}}{\sqrt{|{\rm det}L_{y}|}}
\label{eq:Omega1}
\end{eqnarray}
The expressions for $S_{c}$, $J_{x}$ and $J_{y}$ are given in 
Eqs.~(\ref{eq:Sc1}), (\ref{eq:Jx}) and (\ref{eq:Jy}), respectively.
The conditional probability, 
$P$, is the product of two factors: $P_{0}$, which is described in
the following paragraph, and 
$\Omega^{-1}$. This form is particularly appealing because, as we shall see
below, it is $\Omega$ that accounts for the presence of the competing 
$y$-mode, whereas $P_{0}$ is independent of both $\beta$ and $\gamma$ 
and is therefore insensitive to the presence of $y$. 
To determine $P_{0}$ and $\Omega$ we need to calculate $\det L_{x}$ and 
$\det L_{y}$. 

The calculation of $\det L_{x}$ is straightforward, with the
result that $\det L_{x} \propto \sinh (\alpha T)$. 
Combining this result
with Eq.~(\ref{eq:Sc1}) for $S_{c}$ and Eq.~(\ref{eq:Jx}) for $J_{x}$, 
$P_{0}$ can be written as 
\begin{equation}
P_{0}(X,T)=
\sqrt{\alpha [\coth (\alpha T)-1]}\,\,
e^{\frac{-\alpha X^{2}}{4D} [\coth (\alpha T)-1] }.
\label{eq:P0}
\end{equation}
$P_{0}$ is the conditional probability density that a 
one-dimensional system under the influence of the potential
$-\alpha x^{2}/2$ and Gaussian white noise be located
at $x=X$ at $t=T$ given that it started at $x=0$ at $t=0$. 
As a function of $T$, $P_{0}$ is peaked at a value $T^{*}$ that 
is given by $\coth (\alpha T^{*})=1+2D/(\alpha X^{2})$, 
i.e. 
\begin{equation}
T^{*}=(2\alpha)^{-1}\ln (1+\alpha X^{2}/D).
\label{eq:Tstar}
\end{equation}

The calculation of $\det L_{y}$ 
is not as straightforward; it  
can be expressed as \cite{ref:mckane}
\begin{equation}
\det L_{y}=\frac{h_{2}(T)h_{1}(0)-h_{1}(T)h_{2}(0)}
        {\dot h_{2}(0)h_{1}(0)-\dot h_{1}(0)h_{2}(0)},
\label{eq:detLy}
\end{equation}
where $h_{1}$ and $h_{2}$ are two 
linearly independent solutions of the homogeneous equation $L_{y}h=0$.
The denominator of Eq.~(\ref{eq:detLy}) is the Wronskian of the two solutions.
To evaluate Eq.~(\ref{eq:detLy}), consider the quantity 
\begin{equation}
h_{1}(X,t)=
\exp \left(\beta t-\gamma\int_{0}^{t}\!dt'\,x_{c}(t')^{2}\right).
\label{eq:h1}
\end{equation}  
Taking the second derivative of $h_{1}$ with respect to $t$ we find that 
\begin{equation}
\ddot h_{1}=\Bigl[(\beta -\gamma x_{c}^{2})^{2}
        -2(\dot x_{c})(\gamma x_{c})\Bigr]h_{1}.
\label{eq:ddh1}
\end{equation}
Comparing Eq.~(\ref{eq:ddh1}) with $L_{y}h=0$ from Eq.~(\ref{eq:Lc}), 
we see that if 
$\dot x_{c}=\alpha x_{c}$ then $h_{1}$ is one of the desired solutions. 
Now $\dot x_{c}=\alpha x_{c}\coth(\alpha t)$, so as long as
$\coth(\alpha t)$ is close to $1$, $h_{1}$ is a good solution. 
Recall, however, that for $T<T^{*}$, $P_{0}$ is essentially zero 
[c.f. Eq.~(\ref{eq:P0})]. Thus, it is only values 
of $T>T^{*}$ that are relevant. But $\coth(\alpha T^{*})=
1+{\cal O}(D)$ so indeed we expect that as long as $D\ll 1$, $h_{1}$ 
is a good approximate solution to the homogenous equation $L_{y}h=0$. 
The second linearly independent solution, $h_{2}$ can be 
expressed in terms of $h_{1}$ as 
$h_{2}(t)=h_{1}(t)\int_{0}^{t}dt' \,h_{1}^{-2}(t')$.
With this choice of $h_{2}$, we have that $h_{2}(0)=0$. In addition,
$h_{1}(0)=1$ so that the Wronskian is unity and  
\begin{equation}
\det L_{y}=h_{2}(T)=h_{1}(T)\int_{0}^{T}\!dt\, h_{1}^{-2}(t).
\label{eq:detLy_h}
\end{equation}
Combining this equation with the fact that $J_{y}=h_{1}^{-1/2}(T)$ 
[c.f. Eqs.~(\ref{eq:Jy}) and (\ref{eq:h1})], we find that
\begin{equation}
\Omega (X,T) = h_{1}(X,T)\sqrt{\int_{0}^{T}\!dt\, h_{1}^{-2}(X,t)}.
\label{eq:Omega}
\end{equation}
With this expression for $\Omega$, together with Eq.~(\ref{eq:h1}) for 
$h_{1}(t)$ and Eq.~(\ref{eq:P0}) for $P_{0}$, we have succeeded in deriving
a formula for $P({\bf r}_{x},T|{\bf 0},0)$ that accounts for the Jacobian
prefactor as well as the Gaussian fluctuations about the stationary path
${\bf r}_{c}$. To calculate the analogous formula for 
$P({\bf r}_{y},T|{\bf 0},0)$ [where ${\bf r}_{y}=(0,Y)$], we simply switch
$\alpha$ and $\beta$ and replace $X$ with $Y$.

At this stage we are in a position to calculate, using our expression for 
$P({\bf r}_{x},T|{\bf 0},0)$, the probability that the system flows into
a given well. Currently, we have a simple analytic expression for
$P_{0}$, but to calculate $\Omega$ we need to integrate $h_{1}^{-2}$ over
time. This integration presents no difficulty in principle. However, by making
two approximations we are able to obtain a simple analytic formula 
for $\Omega$. Specifically, we 
approximate the exponential factor 
$\exp[{\frac{\gamma x^2}{\alpha}
      \frac{ \sinh(2 \alpha t)}{2 \sinh(\alpha T)^2}}]$
as $1+\exp[\frac{\gamma x^2}{\alpha} \frac{ \sinh(2 \alpha T) +
                       2 \alpha (t-T) \cosh(2 \alpha T)  }
                       { 2 \sinh(\alpha T)^2 }]$
and omit the contribution from the lower limit ($T=0$) in 
Eq.~(\ref{eq:detLy_h}). This second approximation is due to the fact
that, as explained in the discussion following Eq.~(\ref{eq:ddh1}),
$h_{1}(t)$ is only a good solution for $\alpha t >1$. 
We now obtain the following approximate formula for $\Omega (X,T)$:   
\begin{equation} 
2 \beta \Omega^2 =
\exp\left[ - \frac{ \gamma X^{2}}{\alpha} 
\frac{\sinh (2\alpha T)-2\alpha T}{2\sinh^{2} (\alpha T)}
 \right]  
\left\{ \exp(2 \beta T) - 1 \right\}
\label{eq:Omegaapp}
\end{equation}
We have compared Eq.~(\ref{eq:Omegaapp}) with numerical calculations of
Eq.~(\ref{eq:Omega}), and we find that for the relevant range of parameters 
the results are essentially indistinguishable. 
Thus, Eq.~(\ref{eq:P0}) for $P_{0}$ and
Eq.~(\ref{eq:Omegaapp}) for $\Omega$ together provide an analytic formula
for the conditional probability given in Eq.~(\ref{eq:Pc1}).

We now turn to the computation of the relative probability that the
system flows into an $x$- or $y$-valley. The strategy is to calculate
the total probability flux through the $x$-valleys and the $y$-valleys 
and compare them.  
The probability current, which we denote by ${\bf {\cal J}}({\bf r},T)$, 
is given by
${\bf {\cal J}}\!=\!-P\nabla V \! -\! D\nabla P$, so that the total flux
${\cal F}_{x}$ through an $x$-valley at $X$ is 
${\cal F}_{x}(X)=\int_{0}^{\infty}\!dt\,\int_{-\infty}^{\infty}\!dy
{\cal J}_{x}({\bf r},t)$ and 
the total flux ${\cal F}_{y}$ through a $y$-valley at $Y$ is
${\cal F}_{y}(Y)=\int_{0}^{\infty}\!dt\,\int_{-\infty}^{\infty}\!dy
{\cal J}_{y}({\bf r},t)$, 
where ${\cal J}_{x}$ and ${\cal J}_{y}$ are the $x$- and $y$-components
of ${\bf{\cal J}}$, respectively.    
Denoting by $N_{x}$ the relative probability of flowing into an $x$-valley 
we then have that 
\begin{equation}
N_{x}(X,Y)=\frac{{\cal F}_{x}(X)}{{\cal F}_{x}(X)+{\cal F}_{y}(Y)}
\label{eq:Nx}
\end{equation}

The calculation of the fluxes requires a knowledge of ${\bf{\cal J}}$ 
and hence of $P({\bf r},T|{\bf 0},0)$. In particular, we require this function 
for an arbitrary point in the $x$-valley and not just on the $x$-axis, 
i.e., we require $P({\bf r})$, not simply $P({\bf r}_{x})$. 
The method we have presented may be extended to obtain the full 
functional dependence on ${\bf r}$ \cite{ref:tarlie}, but the 
results given above do not give $P$ any explicit dependence on 
the $y$ variable across the $x$-valley or the $x$ variable across the 
$y$-valley. We will therefore limit ourselves here to showing that, by 
estimating the flux by sampling it on the axis, we can get excellent 
agreement with Monte-Carlo simulations and therefore confirm the 
essential correctness of our approach. A feature of this procedure is
the necessity of fitting $X$ or $Y$. We expect that this will no longer
be required when the flux is calculated, since this should enable 
${\cal F}_{x}(X)$ and ${\cal F}_{y}(Y)$ to be calculated  for large 
$X$ and $Y$ where we would expect them to be insensitive to their actual
values. The only restriction that we will impose on 
$X$ and $Y$ is that they are not too small, for then state-selection
will not have occurred when these points are reached. 
We estimate the minimum value of $X$ to be of 
the order of $X_{min}$, the point at which the force in the $y$-direction 
changes sign. For $V$ given in Eq.~(\ref{eq:V}) we have that 
$X_{min}=\sqrt{\beta/\gamma}$. Likewise, we have that the minimum value of 
$y$ is given by $Y_{min}=\sqrt{\alpha/\gamma}$. 

\begin{figure}
\epsfxsize=\hsize 
\narrowtext
\epsfysize=2.5in
\epsfbox{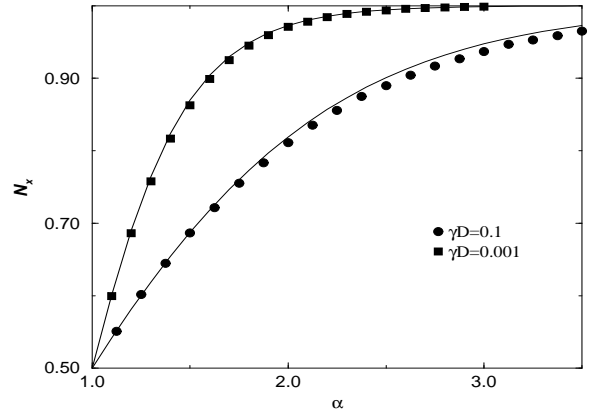}
\vspace{-0.1in}
\caption{Probability of flowing into an $x$-valley as a function of 
$\alpha$. Simulation 
results are for $\beta=1$ and the continuous
curves are our theoretical results.}
\label{fig:results}
\end{figure}

The Monte-Carlo simulations are performed on the Langevin equation 
with $V(x,y)$ given
in Eq.~(\ref{eq:V2}). In Fig.~\ref{fig:results} the results are shown for
a range of values of $\alpha$ and particular choices of $\beta$ and of
$\gamma D$ ($\gamma$ and $D$ always appear in this combination, since the
effect of the interaction is to renormalize the noise). 
The theory we have outlined here is seen
to be in excellent agreement with the simulations. For $\gamma D\!=\!0.1$ 
we have taken $X\!=\!X_{min}$ and $Y\!=\!Y_{min}$ and for 
$\gamma D\!=\!0.001$ we have taken $X\!=\!1.83 X_{min}$ and 
$Y\!=\!1.83 Y_{min}$. Comparison for other values
of the parameters, a determination of the region of validity of our 
approximation in parameter space and further improvements of the method will 
also be discussed elsewhere \cite{ref:tarlie}.

In this Letter we have presented a systematic method for determining state
selection from an unstable stationary state, when multiple metastable states 
compete for occupation. Previous methods have not addressed this question
directly. Our treatment has the added advantage of
yielding closed form, analytic expressions for the
conditional probability distribution. Finally, we emphasize that, although
we have focussed on a specific potential system with two degrees of freedom 
for illustrative purposes, our theory is neither restricted to potential 
problems nor to systems with only two degrees of freedom.

\vspace{-0.15in}

We thank Ken Elder for useful discussions, and the Universities of 
Chicago and Manchester for hospitality. 
This work was supported in part by EPSRC grant GR/K79307 (AJM) and
by the NSF (DMR-9415604) (MBT).

\end{multicols}

\begin{references}
\bibitem{ref:quasi} By quasi-one-dimensional we mean the potential is 
spherically symmetric and the system is therefore effectively one-dimensional.
\end{references}
\end{document}